# Practice papers

# Understanding and managing blockchain protocol risks

## Alex Nathan
Head of Analytics and Co-founder, Metrika Inc.

Alex Nathan is co-founder and Head of Analytics at Metrika. Over the last four years, Alex has been analysing large amounts of data to understand the performance, reliability and operational health of blockchain networks. Alex holds a PhD in industrial engineering from Northwestern University, where he researched applied and methodological aspects of large-scale machine learning.

Metrika Inc.
E-mail: alex@metrika.co

## Dimosthenis Kaponis
Chief Technology Officer, Metrika Inc.

Dimosthenis is currently Chief Technology Officer (CTO) at Metrika, overseeing the company's technology vision and strategy. He was previously CTO of Netdata, an infrastructure monitoring company and Beat, a Daimler/BMW subsidiary in the ride hailing space. Starting his professional software engineering career while still in high school and working in parallel to his studies, he dropped out of his PhD to found Cosmical Technology, a boutique software/consulting company, at the age of 25. At Cosmical he co-created two of Greece's popular smartphone apps of the early 2010s, AthensBook and ThessBook. He holds an M.Eng from Imperial College London.

Metrika Inc.
E-mail: dimosthenis@metrika.co

## Saul Lustgarten
Chief Product Officer, Metrika Inc.

Saul Lustgarten is the Chief Product Officer at Metrika. Before Metrika, he was Head of Product at Tulip Interfaces, and co-founder of Aprovecha.com and Ringtu. He holds an MBA from MIT Sloan School of Management and a BSc in applied mathematics, economics and philosophy from Brown University.

Metrika Inc.
E-mail: saul@metrika.co

**Abstract**   This paper addresses the issue of blockchain protocol risks, a foundational category of risks affecting Distributed Ledger Technology (DLT) which underpins digital assets, smart contracts, and decentralised applications. It presents a comprehensive risk management framework developed in collaboration with financial institutions, blockchain development teams and regulators that applies a traditional risk management taxonomy to address certain overlooked blockchain protocol risks. The approach offers a structured way to identify, measure, monitor and report blockchain protocol risks. The paper provides real-world use cases to demonstrate the practicality and implementation of the proposed framework. The findings of this work contribute to the evolving understanding of blockchain protocol risks and provide valuable insights on how these risks affect the adoption of DLT by financial institutions.

## INTRODUCTION — WHAT ARE BLOCKCHAIN PROTOCOL RISKS?

Blockchain is a distributed ledger technology (DLT) that enables secure, transparent and tamper-proof recording and sharing of data without the need for a centralised authority or intermediary.[1] Blockchain networks consist of nodes, each holding a local copy of the recording mechanism or ledger, which work together to validate new transactions via consensus algorithms. Depending on their design and functionality, blockchain networks can support various use cases, ranging from near-instant, peer-to-peer value transfers to the execution of smart contracts containing more complex logic.

Through collaborations with risk practitioners, blockchain developers and regulators, it became apparent that their primary concerns regarding public blockchain risk revolve around regulatory, cybersecurity and financial aspects. Specifically, the identification and mitigation of anti-money laundering (AML), 'know your customer' (KYC), 'know your transaction' (KYT) and sanctions risks (the 'tip of the iceberg', as shown in Figure 1), which have more defined regulations, were prioritised along with smart contract vulnerabilities and financial risks related to custody, liquidity or price volatility. While the importance of these risks cannot be understated, the attention of this paper is directed to a frequently overlooked set of risks: blockchain protocol risks.[2]

The motivation for this work stems from the fact that, despite their resilient design, blockchain networks can malfunction due to software bugs, attacks, or experience downtime or performance degradation due to mismanaged nodes. An additional reason for focusing on blockchain protocol risks is proposed legislation, such as the Digital Operational Resilience Act (DORA) in the EU[3] and the McHenry–Thompson bill[4] on market structure in the US, which may require financial institutions to develop comprehensive capabilities for business continuity, risk management and reporting for all information and communication technology matters. Blockchain protocol risks herein refer to all risks threatening the operational health and viability of a blockchain network.

Protocol risk differs from AML and sanctions risk in that the latter typically pertains to a specific individual transaction or entity. Similarly, blockchain risk is distinct from the cybersecurity risk associated with a smart contract or decentralised app (dApp), as the latter is mostly[5] confined to that corresponding smart contract or dApp, whereas the former potentially affects the system as a whole. That is, blockchain risks affect all dApps, smart contracts, and transactions on top of a given blockchain network: in the event of a catastrophic incident, all applications, assets and smart contracts on top of the affected blockchain system would be adversely affected as well. Notwithstanding the foregoing, protocol and regulatory risks are not mutually exclusive; as discussed below, there are some regulatory considerations when assessing blockchain risks.

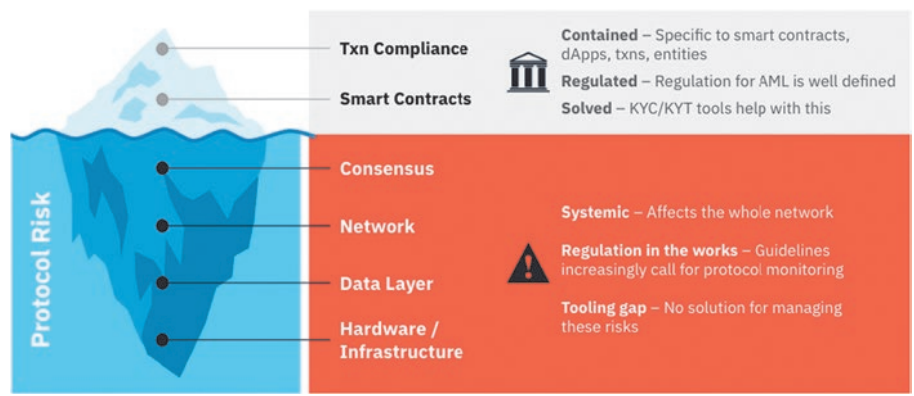

**Figure 1:** Blockchain risk iceberg

The remainder of this paper is organised as follows: the next part introduces the proposed risk management framework, which is followed by an overview of identifying and measuring key risk indicators (KRIs). The paper concludes with some real-world examples showcasing the implementation of the framework in practice.

## A FRAMEWORK FOR MANAGING BLOCKCHAIN PROTOCOL RISKS

Blockchain protocols and their resulting networks vary across several dimensions, such as their consensus mechanisms, the transaction types they support, the centralisation of their validators, their scalability, smart contract environment, accounting methods (UTXO versus account-based[6]) and even the ethos of their user communities. A useful analogy for the use of these networks by financial services companies is to view different blockchain protocols (eg Bitcoin, Ethereum, Solana, Algorand) as distinct technology software vendors, each with their own characteristics. When considering building a product or a service on a blockchain network, it is important to evaluate all available 'vendors', based on both their features and risk profiles.

The framework proposed in this paper enables risk functions to identify and mitigate new, emerging or evolving risks associated with blockchain technology. Moreover, it facilitates both prospective evaluations of blockchain networks and ongoing monitoring in their use by financial institutions (Figure 2).

At a high level, the purpose of each step is listed below.

1. *Identify and measure*: Identify the blockchain risks affecting the business case at hand and quantify KRIs. These KRIs should be mapped to monitorable on-chain and off-chain metrics.
2. *Manage*: Define the appropriate thresholds or tolerance bands for each KRI, which help determine when aspects of the blockchain network would be considered 'at risk' or at 'unhealthy' levels. These thresholds can be either static or dynamic (eg moving averages). More advanced alert capabilities can also be useful, whereby an end user can define trend-based alerts, anomaly-detection-based alerts, and alerts that depend on the correlation between two or more KRIs.
3. *Monitor*: Track KRIs through real-time dashboards and APIs. Predictive and real-time alerts should be triggered and routed to the appropriate communications channels. Periodically adjusting thresholds and retraining of predictive models may be necessary depending on the business risk appetite, network maturity and applicable regulatory frameworks.
4. *Report*: Maintain records of KRIs and risk mitigation responses for streamlining audits, regulatory approvals and for proof of compliance.

The continuous monitoring of blockchain networks is an iterative process that may lead to identifying new risks, which may require expanding or adjusting the established KRIs. This paper primarily focuses on the 'identify and measure' step, which requires a deep understanding of blockchain technology. In Appendix A, a short real-world example is provided illustrating the 'managing' and 'monitoring' steps.

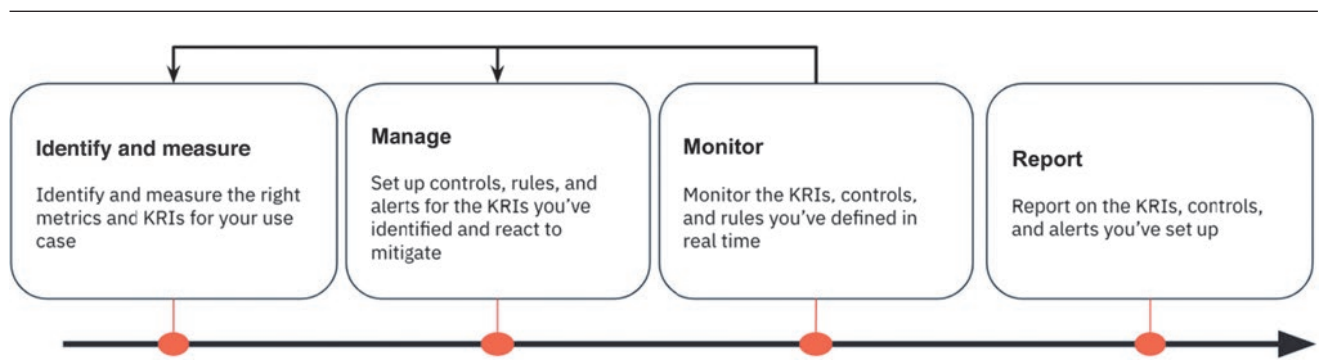

**Figure 2:** Risk management framework

# IDENTIFY AND MEASURE: DECOMPOSING BLOCKCHAIN PROTOCOL RISKS

The risk taxonomy proposed in this paper groups blockchain protocol risks into the following categories: centralisation; network reliability and performance; security; financial; people; and regulatory. For each category, illustrative examples are provided of relevant risks and KRIs that can help risk practitioners monitor and mitigate blockchain risk.

## Obtaining the data

Obtaining the necessary data is a crucial prerequisite for implementing the proposed framework in practice. In centralised systems, such as traditional software-as-a-service vendors, the responsibility for measuring and managing these risks falls on the vendor. However, in decentralised blockchain systems, the organisations that utilise the technology are responsible for collecting and analysing the data and managing the risks uncovered by it.

While there is no single data source/solution that encompasses the wide range of data for all blockchain networks, some key sources include:

- data from blockchain nodes;
  - APIs exposed by blockchain nodes, which are the single source of on-chain data (eg transactions);
  - logs produced by blockchain nodes, which can contain useful off-chain information (eg arrival times of votes and proposals);
  - machine level metrics, such as CPU and memory usage, if collected at scale, can help assess and predict the overall health of a network.
- the networking layer of any blockchain protocol allows nodes to identify and exchange with each other. Tapping into peer-to-peer networks can provide a wealth of information, such as the total number of nodes in the network and IP addresses of each identified node;

- data from public code repositories, which can help track developer activity;
- monitoring social media feeds for sentiment analysis;
- monitoring government websites in real-time for news on legal actions and enforcements.
- partnerships with ecosystem participants, which can unlock data that may initially seem difficult to collect (eg mapping between validators and entities/organisations).

## Dealing with missing data

The handling of missing, incomplete, or erroneous data significantly influences risk quantification and necessitates the use of various strategies while understanding their limitations and potential biases. Techniques such as stress testing, scenario analysis and data collection enhancement are crucial. For risk categories such as financial data, statistical methods of interpolation or maximum likelihood estimation can be used. For others, such as network performance data, Bayesian inference may help, albeit with additional assumptions. Last, transparency about the methods and their potential impact on risk estimates is crucial in decision-making contexts.

# RISK CATEGORIES AND KRIs
## Centralisation

Decentralisation is a fundamental principle of blockchain networks, safeguarding their security, transparency and trustworthiness. Blockchain systems, however, are susceptible to various attack vectors that can lead to a single entity or group of entities controlling the majority of the network, enabling them to manipulate transactions, double-spend tokens, and potentially compromise a protocol's integrity. Measuring and tracking centralisation using KRIs is crucial to maintaining network resilience.

- *Network centralisation*: Measuring the overall decentralisation of a blockchain network (either in terms of stake for Proof of Stake [PoS] networks or hash power for Proof of Work [PoW] networks) requires a mapping of nodes or miners/validators and their owners/operators, as it is possible for a single entity to operate a large number of nodes or validators. For example, as of this writing, blockchain.com reports that the Foundry USA mining pool controls 28.765 per cent of Bitcoin's hash power.[7] Using a reliable mapping, key performance indicators (KPIs) such as the Nakamoto Coefficient (51 per cent+ control of the network by any one or group of entities) can be used to monitor the decentralisation of a network. Other popular decentralisation metrics include the Gini coefficient[8] and the Theil index.[9]

- *Cloud service provider centralisation*: Centralisation via cloud providers is a frequently overlooked risk. If all nodes or validators of a blockchain system are hosted on the same cloud provider, an outage could jeopardise the network. The Solana–Hetzner incident in November 2022 exemplifies this risk; the German cloud service provider blocked server access to Solana validators, taking over 1,000 validators offline.[10] Reports suggested that around 40 per cent of Solana validators, equating to 20 per cent of the network's stake, were hosted by Hetzner at the time.

- *Geographic centralisation*: A geographically distributed infrastructure ensures resilience to natural disasters, geopolitical, jurisdiction and regulatory risks. In PoS networks, a potential centralising force for geographic concentration is latency: node operators may want to be geographically close to most other nodes to receive transactions as fast as possible. In PoW networks, where block mining requires significant energy consumption, it is likely to find clusters of nodes concentrated in areas where electricity costs are low.

- *Block builder centralisation*: In blockchain networks with out-of-protocol proposer builder separation (eg Ethereum), centralised block-building via exclusive order flows can lead to transaction censorship; essentially, block builders could manipulate the market and exclude transactions originating from one source in favour of another. To assess network censorship, one must compare and contrast 'on-chain' transactions with transactions in the Mempool (a waiting area for transactions submitted to the network for consideration). All other things being equal, a wide discrepancy between Mempool and on-chain transactions can indicate whether specific block builders are censoring transactions.

## Network reliability and performance

Traditional technology vendors offer their clients service-level agreements (SLAs), which specify penalties and remedies if the SLA is broken. In decentralised systems, there is no central authority that can offer an SLA, which means that the burden of reliability and performance monitoring falls on the ecosystem participants. In theory, decentralised infrastructures should be more resilient than their centralised counterparts because they lack single points of failure. In practice, however, there have been numerous incidents (eg Solana[11], Avalanche[12]) where protocols have suffered from degraded performance or a full outage. Monitoring the performance and reliability of a decentralised network is challenging in the absence of a centralised monitoring system. As such, evaluating the overall state of the network requires piecing together the puzzle by analysing on-chain and off-chain signals. A non-exhaustive list of KRIs that can help monitor the overall health of the network is provided below.

- *Consensus performance*: Nodes or validators participating in the consensus mechanism of a protocol typically perform two crucial duties: (1) propose new blocks and (2) vote on the validity of proposed blocks in a timely manner. If validators stop performing their duties as expected, the network may eventually fail. For example, in certain protocols, if validators stop casting their votes the network may stall, as no new blocks will receive the necessary majority of votes to be recorded on-chain. The exact consensus performance metrics may differ between protocols, but typical KRIs for consensus performance include the number of missed block proposals by validators, timeliness of vote arrival, and the number of voting validators divided by total validators. These same measurements can also assess individual validator performance. Another useful metric that is a by-product of validator participation is transaction finality. Finality guarantees that transactions are permanently stored on-chain. Metrics such as time to finality and distance between the current and latest finalised block are accurate indicators of consensus health and performance.

- *Network liveness*: Two important metrics can help assess a network's liveliness, ie its ability to continuously process new transactions: (1) transaction throughput, and (2) block generation/production rate. Transaction throughput is typically analysed through the Transactions Per Second (TPS) metric. When evaluating the liveness of various networks, it is crucial to take a holistic approach instead of solely focusing on a single KRI. Comparing networks based on metrics such as TPS alone can result in inconsistencies.

- *Network upgrade readiness*: Blockchain networks, still in their early stages, frequently undergo software updates for improved functionality, performance and security. Upgrading in a decentralised context is challenging and requires coordinated action from disparate parties. Network readiness can be assessed by the proportion of nodes/validators using the latest client software version in advance of an imminent upgrade. During major upgrades, or hard forks, outdated nodes cannot participate in the network, and a consensus stall may occur if participants do not upgrade on time.

## Security

When companies select traditional vendors to provide products or services, certain due diligence documentation and evidence are requested, such as completed audits, penetration testing and ISO and SOC2 certifications, to ensure that the vendor maintains adequate governance and control of their information security programme risks. Although there are no equivalents to these certifications for blockchain systems yet, the ability to identify security vulnerabilities has several advantages over traditional technologies. Some example KRIs for monitoring the security of a blockchain network are provided below (other cybersecurity risks such as smart contract security[13] are important, but beyond the scope of this paper).

- *Economic security*: This refers to the security resulting from built-in incentives in the tokenomics of the blockchain.
  - In a PoW blockchain, miners compete using computational power to solve a complex problem, enabling them to add new blocks to the chain. For an attacker to gain network control, they would need to own a majority of the network's computational power or hashrate. However, due to the prohibitive costs of acquiring and maintaining this majority hashrate, attacks are generally considered economically unfeasible.
  - In a PoS blockchain, it is possible to track total value staked (TVS) as a measure of economic security because validators must stake assets as collateral to participate in the consensus mechanism. This means that the higher the TVS, the more costly it becomes to carry out a successful attack. For example, if an attacker needed to control 51 per cent of the network to perform an attack on Ethereum Mainnet, it would require over US$13.5bn as of the publishing of this paper. Additionally, if the attacker were successful, they would risk losing their staked assets since any malicious activities lead to slashing (where the stake is automatically forfeited).

- *Security ratio*: Related to TVS, this is a ratio of the total value secured by a protocol over the economic security of that protocol. Put another way, the value of all the assets secured by a blockchain, over the value of securing the blockchain itself (stake in PoS protocols). A lower TVS ratio indicates a more secure chain, as an attack would be more expensive to carry out. Conversely, a higher security ratio may provide an attractive opportunity for attackers, since the cost of an attack would be relatively low compared to the rewards to be gained.

- *Public code audits*: Public blockchain networks are powered by open-source code, which means that the code can be inspected and audited for security vulnerabilities by anyone. Cybersecurity firms and individuals regularly conduct audits of blockchain codebases and share their findings publicly, providing assurance for reliance on the protocol.

- *Bug bounty programmes*: Some blockchain networks have open bug bounty programmes that effectively crowdsource penetration testing and other types of quality/security checks from the community. Programmes and results are often public, so they can be used to assess the resiliency of the protocol and indicate how much the core development team invests in security. Other attributes of bounty programmes, such as reward size, are also important.

## Financial

Financial risks of traditional vendors are associated with the potential financial losses and instability of the business as a result of the environment in which they operate. Factors such as capitalisation, funding, revenue and profitability are useful metrics to assess the longevity of a business. A similar analysis can be applied to blockchain networks.

- *Market capitalisation, circulating supply*: A token's market capitalisation and circulating supply are commonly used metrics to judge the expected stability, overall value and potential longevity of a project. If analyses suggest market inefficiencies, this may be a precursor to financial volatility.
- *Fundamentals*: An alternative approach for evaluating blockchain projects is by analysing the revenues and costs of blockchain networks. Protocols sell blockspace as users must pay a fee for their transaction to be included in a block. The costs for blockchain networks are the rewards provided to miners or stakers to secure the chain and ensure consensus is reached. To evaluate the value of a blockchain, it is possible to use a similar approach to valuing a company by estimating future payments through discounted cash flows, taking into account the expected growth rate of the protocol.
- *Validator economics*: Node operators provide a valuable service to the blockchain in that they validate transactions and add them to the blockchain (in PoW protocols, this process is called mining, in PoS, it is called staking). In return for their service, they earn rewards.

However, running a node is not free. Therefore, when evaluating the financial viability of the network, it is important to consider validator economics. If operating validators is not profitable or does not provide competitive returns, this can be a leading indicator of network unsustainability.

- *Tokenomics*: This refers to the incentive mechanism, monetary policy and initial distribution of a blockchain's native currency. Understanding the tokenomics of a protocol is important for understanding the value of the underlying digital asset, but also to understand the incentives for the network. Part of a blockchain's tokenomics also includes the token distribution and incentive structure of the team behind the protocol. Most blockchain networks are fairly transparent about token distribution and vesting periods.
- *Treasury and proof of reserves*: Blockchain development teams generally maintain their protocol treasuries on-chain so anyone can check their balance and spending rules. This helps mitigate the risk of blockchain projects not having enough resources to meet their payment obligations (eg software engineers developing the blockchain codebase).

## People

As with any project, blockchain systems rely on the people involved, which means they are exposed to common people-related risks such as turnover, attrition and succession planning. Evaluating these risks with blockchain teams in the absence of an HR department is a non-trivial pursuit, although possible nonetheless. Below are some KRIs that can be used.

- *Core developer team size and engagement*: Since protocols are open source, it is possible to look at the contributions in public code repositories to see who is behind them and track if the core development team is growing. Additionally, it is possible to measure activity and engagement through metrics such as number of commits and new lines of code over a period of time.

- *Developer community*: Similarly, a growing developer community can be a positive sign of a network's potential, as more people are able to build products and services on top of the protocol.
- *Key person dependencies*: Blockchain systems are not immune to the key person dependency. Some networks are more susceptible to this risk than others. For example, no one knows who was behind Bitcoin, as the author of the whitepaper, Satoshi Nakamoto, remains anonymous. Others, such as Ethereum, however, do have influential personalities behind them.
- *Social media sentiment*: Successful protocols have healthy online communities. A protocol's online user presence can be measured to see if it is growing across communications channels (eg Discord, Twitter, Reddit), and whether the overall sentiment of the community's members is positive.

## Regulatory

Regulation poses significant risks to blockchain networks, as legislative and regulatory decisions can greatly affect a network's viability. For instance, if a specific blockchain's native asset is classified as either a commodity, security or other asset type, it may be subject to a different set of regulatory requirements, potentially affecting the network's appeal to developers and users. Some KRIs to monitor include the following:

- *Legal risk*: The likelihood of negative consequences resulting from uncertainties or ambiguities in the legal and regulatory environment surrounding blockchain technology and its applications. Factors that may affect financial institutions using blockchain technology would include legal disputes, violations of law, fines, various legal opinions, case law and potential legal liability issues. For example, several notable legal developments (eg SEC *v* Ripple[14] and Kraken's Settlement with the SEC[15]) in the US have the ability to change the operating environment and parameters of use for US-based institutions leveraging public blockchain networks. A relevant KRI is the number of pending lawsuits or other legal actions against a legal entity developing a blockchain network.
- *Political/jurisdiction risk*: Related to the legal risk described above, this risk refers to the likelihood of political decisions, events, or conditions that would significantly affect the stability and operating environment of a country or region with regard to financial institutions and their use of public blockchain systems. This relates to the geographic centralisation risk as described earlier, as key components of infrastructure or a significant part of the network in a given jurisdiction may be compromised due to impacts from regulation, policy changes, political instability, regime changes, corruption, sanctions or trade barriers. If the network is more decentralised, it is less likely to be adversely affected by individual jurisdictions.
- *Transaction and block building compliance*: Tracking sanctioned persons and addresses across different jurisdictions (eg OFAC's SDN list) is important to avoid interacting either directly or indirectly with sanctioned individuals or entities. Furthermore, in protocols such as Ethereum, where there is proposer–builder separation, validators may be forced to make a choice: sell their block space to regulated/compliant block builders or not. While there are currently several OFAC-compliant block builders available for validators to choose from on the Ethereum network, it is important to note that entities that self-identify as regulated or 'OFAC-compliant' may not always be reliable or accurate in their claims.[16]
- *Environmental impact*: Environmental, social and governance (ESG) refers to a set of standards to which enterprises and financial institutions have become increasingly more attuned for a variety of reasons. The EU, for instance, has introduced the Corporate Sustainability Reporting Directive (CSRD), which aims to enforce environmental and social reporting for large companies. To track the environmental impact of their blockchain activities, companies can track metrics such as energy consumption from the hardware used in mining and staking.

# EVALUATING AND QUANTIFYING RISK

Upon finalising the desired list of KRIs, financial institutions may wish to aggregate these metrics to provide a portfolio view of risks via quantitative risk measures/scoring. This composite score can encapsulate the various facets to portray the totality of risk within an organisation. In this section two alternative evaluation functions are briefly discussed.

## KRI-based weighting

Utilising a weighted sum of KRIs is an effective methodology for quantifying the holistic risk profile of an organisation. The weight coefficients assigned to each KRI should reflect both intrinsic and extrinsic organisational attributes. Intrinsic attributes can include an organisation's risk appetite, the cruciality and impact of each dimension, the effectiveness of internal risk mitigation measures, and the nature of its operations. For example, financial institutions may prioritise financial and regulatory KRIs, whereas tech–based enterprises might focus more on network performance risk. Extrinsic factors, on the other hand, include prevailing regulatory standards, market conditions and industry–specific standards. Taking all this information into account, organisations can assign lower weights KRIs they consider as low risk and *vice versa*:

$$R = \sum_i w_i x_i$$

where $R$ is the risk score at any given moment in time, $w_i$ = weight for KRI $i$, $x_i$ = the value of KRI $i$ ($x_i$s would need to be normalised). With this approach, the organisation will need to provide the normalised weight, $w_i$, for each KRI employed.

## Category-based weighting

Given the possibly very large number of KRIs involved, second-order weighted sums can be employed as an efficient tool to capture the intricate relationships across risk categories, providing a more nuanced representation of an organisation's risk landscape.

$$R = \sum_c \left[ W_c \times \sum_i (w_{ci} \times x_{ci}) \right]$$

In this approach, each risk category is represented as a weighted sum of its respective KRIs. A key assumption that makes this method considerably easier to implement than the KRI-based one is that the first-order weights of individual KRIs ($w_{ci}$) are readily available.[17] These template weights can either be obtained with the help of blockchain experts, or could follow from widely accepted industry standards. Consequently, instead of assigning weights to individual KRIs individually, organisations can assign weights to the risk categories ($W_c$) as a whole. These risk categories can, in turn, mirror more accurately an organisation's strategic priorities, risk appetite and specific vulnerabilities.

# APPLYING THE FRAMEWORK: USE CASES

This section demonstrates the implementation of the proposed framework in real-world enterprise activities involving blockchain technology. It starts by presenting the use cases, and subsequently focuses on certain KRIs that institutions may monitor or predict, along with their associated controls and risk mitigation measures. The use cases discussed may be susceptible to risks beyond blockchain level risks, but these are not addressed by this work.

## Custody

Digital asset custody broadly refers to the storage and protection of digital assets on behalf of their owners. While similar in spirit to traditional asset custody, digital asset custody involves securely storing the cryptographic keys of their owners, thereby storing the assets themselves. Digital asset custodians typically offer additional services to their clients, including the ability to buy and sell digital assets. Institutions and enterprises maintain secure custodial solutions, either through software solutions that enable self-custody or by using trusted third-party custodians for their digital assets. These custodians offer enterprise-grade security, such as multi-party computation (MPC), which can be used with cold, warm and hot wallets. There are several digital asset custody providers, ranging from cryptocurrency exchanges/custodians (eg Coinbase), financial institutions (eg BNY Mellon[18]) and specialised custody providers (eg Fireblocks). A recent survey[19] sponsored by BNY Mellon indicated that 41 per cent of institutional investors currently hold cryptocurrency in their portfolios, and 66 per cent of respondents claimed they would increase their digital asset activity if services such as custody and execution were available from trusted institutions.

## Staking

Staking is the mechanism by which individuals and organisations can participate in the consensus mechanism of a PoS blockchain network and earn rewards for completing their duties. Highlighted below are the two most popular options for financial institutions to take part in the staking business, each of which has its own risks and benefits[20]:

1. *Running a validator*: When organisations run their own validators they are completely responsible for hosting and managing the underlying infrastructure. In order to run a validator, organisations must have the technical knowledge and expertise to set up and maintain the hardware, client software, network, validator keys and many other aspects to be successful.
2. *Use a staking as a service provider*: Organisations can use staking as a service (SaaS) providers to operate dedicated validators on their behalf. These providers handle the configuration and maintenance of the validators, but the organisation will pay a fee, typically a percentage of the rewards earned by the validators.

In 2021, J. P. Morgan analysts forecasted that staking yields across the blockchain industry could reach US$40bn by 2025.[21]

## Trading

Trading digital assets involves buying and selling cryptocurrencies or other digital tokens on various online platforms. Unlike traditional trading, cryptocurrency markets operate 24/7, are highly volatile and generally are less regulated. Cryptocurrency traders can use a variety of strategies, including but not limited to:

- *Scalping*: scalping can be a day or high-frequency trading technique that involves buying and

selling individual assets multiple times. Scalping typically requires the rapid entry and exit of positions, which highlights the relevance of the underlying blockchain network's performance and reliability in facilitating such transactions.
- *Arbitrage*: arbitrage refers to taking advantage of market inefficiencies for profit. Arbitrage in the blockchain world exists in varieties such as cross-asset/triangular arbitrage, cross-border arbitrage and cross-chain arbitrage. Similar to scalping, arbitrage opportunities require fast execution, as they tend to be very short-lived. Cross-chain arbitrage strategies involving asset transfers via bridges[22] have even more implications for risk management practitioners, as they require the risks to be managed across multiple blockchain networks.

## Centralised Exchanges

Centralised cryptocurrency exchanges are centrally controlled online platform marketplaces where users can buy, sell and trade cryptocurrencies with other users. These orders are matched with other users willing to buy or sell at the same price. Once a match is found, exchanges facilitate the transaction and transfer the cryptocurrency to the buyer's account. Exchanges typically perform this matching 'off-chain', to make transactions more cost efficient and faster for their end users. Exchanges are exposed to a wide range of risks, ranging from hacks[23] to extensive regulatory scrutiny.[24]

## Tokenisation

Tokenisation, creating digital representation ('digital twins') of assets such as bonds, stocks and real-estate on a blockchain, is one of the most exciting use cases of blockchain technology for financial institutions. The benefits of tokenisation for financial firms range from cost savings when creating, issuing, managing or tracking ownership of assets to access to a more inclusive range of investors as a result of fractional ownership. The BNY Mellon survey mentioned previously found that 91 per cent of institutional investors are interested in tokenised products. Example risks when tokenising assets in public blockchain networks include scalability, stability, security, privacy and potential gaps in existing regulatory frameworks.[25]

## Applying the framework

Table 1 presents some of the KRIs that an institution may choose to monitor along with controls and actions to manage the associated risks. Some of these KRIs and how they affect different use cases are shown below.

Continuous monitoring of KPIs enable quick detection of deviations and provide the ability to mitigate risks in a timely manner. As previously noted, due to the 24/7 operation of blockchain networks, crucial incidents that necessitate immediate responses and interventions can occur at any moment and their duration will be unknown. It should also be noted that all remediation actions may need to be logged to fulfil reporting and compliance obligations. Appendix A shows how

**Table 1:** Example KRIs across five use cases: custody, exchanges, staking, tokenisation and trading

| Use cases | Identify and measure | | | Manage |
|---|---|---|---|---|
| | Risk category | Risk subcategory | Example KRI | Alerting conditions[26] |
| Custody, exchanges, staking, tokenisation, trading | Decentralisation | Network decentralisation | Nakamoto Coefficient | $<= 20\%$ |
| Custody, exchanges, staking, tokenisation, trading | Decentralisation | Cloud provider decentralisation | Max % of nodes hosted by single provider | $>= 20\%$ |
| Custody, exchanges, staking, tokenisation, trading | Decentralisation | Geographic decentralisation | % nodes in one country | $>= 20\%$ |
| Staking | Network reliability and performance | Consensus performance (own validators) | 1. % of successful votes 2. Consecutive missed proposals | $=< 95\%$ $> 1$ |
| Custody, exchanges, staking, tokenisation, trading | Network reliability and performance | Consensus performance (overall network) | 1. % of successful votes 2. Consecutive missed proposals | $< 95\%$ $> 4$ |
| Custody, exchanges staking, trading, tokenisation | Network Reliability and Performance | Network Upgrade Readiness | % of nodes/validators in latest software version | $< 50\%$ |
| Custody, exchanges, staking, trading, tokenisation | Network Reliability and Performance | Network Liveness | Transactions per second (TPS) | $< 100$ |
| Custody, exchanges, staking, trading | Financial | Market capitalisation | Hourly % change in market capitalisation | $> 5\%$ |
| Staking, trading | Financial | Validator economics | APY/operating costs | $< 30\%$ |
| Custody, exchanges, staking, trading, tokenisation | Security | Economic security | TVS | $< \$10B$ |
| Custody, exchanges, staking, trading, tokenisation | Regulatory | Legal risk | Pending lawsuits against legal entity/developer | $>= 1$ |
| Custody, exchanges, trading, tokenisation | Regulatory | Transaction compliance | % of non-compliant OFAC transactions | $> 1\%$ |
| Staking | Regulatory | Block building compliance | % of non-compliant OFAC blocks | $> 1\%$ |

institutions can leverage risk management platforms to operationalise this framework.

1. KRI: Maximum percentage of nodes hosted by a single provider

The Solana–Hetzner incident is a great reminder of the perils of cloud provider centralisation. A high value for this metric can be a cause of concern, as an outage or some other catastrophic event at the cloud provider level could potentially make the network susceptible to attacks (this is also true for other decentralisation KRIs):

- *Custody*: In the event of an attack, custodians could encounter difficulties carrying out client instructions, such as wallet rebalancing — rebalancing is the act of moving assets from cold to hot wallets and *vice versa*. Timely communication with asset owners can prove useful. Regulated digital asset custodians may also need to keep track of all actions taken during such an incident for reporting and compliance purposes.
- *Trading*: If the network is indeed attacked, traders could have their on-chain transactions censored and/or reverted, which could lead to financial losses. Some potential actions for traders include closing positions on the affected network and suspending all trading.
- *Staking*: Validators can be affected by cloud provider outages as this translates to missed revenue and, depending on the underlying protocol, accrual of penalties for non-participation. Validators have the power to mitigate this risk by deploying their nodes in a more decentralised manner. Furthermore, they can raise awareness of the issue by contacting other network participants (eg partners and clients) and encourage them to take action.
- *Exchanges*: Centralised exchanges have historically been the targets of attacks. For example, in 2020, OKEx lost ~US$5.6m in Ethereum Classic in two 51 per cent attacks (19). If an attack seems plausible, exchanges can suspend deposits and withdrawals activity for a particular token or increase transaction confirmation time requirements.[27]

- *Tokenisation*: A severe network incident would impede access to the assets and the ability to verify and transfer ownership of those assets.

2. KRI: Percentage of successful votes

As mentioned above, if validators fail to cast their votes on the validity of new blocks, the network can lose its ability to finalise transactions or even stall. Such an incident would have implications across our use cases:

- *Custody*: Custodians may face challenges in processing transactions or executing client instructions. If a consensus stall appears inevitable, custodians may proactively communicate with their clients and devise contingency plans.
- *Trading*: Traders rely on the ability to buy and sell assets on time. A stall can lead to delayed or failed trades, resulting in financial losses or missed opportunities. To mitigate their risks, traders could halt all trading activities.
- *Staking*: Stakers earn rewards by proposing new blocks and voting on the validity of proposed blocks. A consensus stall can affect stakers' ability to earn rewards, and, in certain networks, it might also lead to financial penalties. Stakers with advanced predictive capabilities could withdraw staked tokens ahead of time if a catastrophic event is deemed imminent.
- *Exchanges*: Similar to trading and custody use cases, exchanges may face issues with withdrawals and deposits. Furthermore, given that exchanges perform their order matching off-chain, exchanges may have to halt all trading activities for assets on the affected chain to prevent price discrepancies.
- *Tokenisation*: Similar to previous use cases, a stall would impede the ability to mint or transfer tokens.

3. KRI: Percentage of nodes/validators in the latest software version

Network upgrades are an integral part of blockchain systems. In an ideal scenario, most of a network should be ready for an upgrade ahead of the planned date; however, this might not always be the case. Whether it is because a hard fork itself is contentious, or the various network participants simply fail to act promptly, the protocol may be at risk:

- *Custody*: Custodians typically have fork policies for their clients to review. If a failed network upgrade seems likely, proactive communication can help safeguard assets in custody.
- *Trading*: Anticipation of an unsuccessful upgrade may lead to a decline in trading volume. During periods of uncertainty, traders may consider more conservative strategies or even halt trading activities.
- *Staking*: Similar to the previous example, stakers may be unable to earn their rewards, and, depending on the network, they may be facing additional financial losses. Withdrawals are always an option for stakers to protect their interests. Other options include outreach to other network participants to raise awareness about the upcoming upgrade.
- *Exchanges*: Exchanges tend to operate validators as well, but, for them, actions like unstaking may not be feasible; therefore, the option of raising awareness about an upgrade may be more appropriate. If, despite their efforts, the failed upgrade appears imminent, a decision must be reached with regard to on-chain and off-chain trading of any affected assets.
- *Tokenisation*: In the event of a hard fork where a subset of the validators does not upgrade in a timely manner, the tokenised assets would effectively be duplicated across the two forks, potentially affecting the utility and value of these assets.

4. KRI: Percentage of non-compliant OFAC blocks and transactions

OFAC regulations apply to US persons, but similar sanctions might be relevant in multiple jurisdictions. While OFAC compliance has been top of mind for custodians and exchanges since the early blockchain days, the recent events with Tornado Cash landing on the SDN list[28] have brought it back to the spotlight.

- *Custody*: Custodians are required to screen clients against the OFAC sanctions list to ensure compliance.
- *Trading*: Traders should have the proper mechanisms in place to avoid transacting with individuals or entities on the OFAC sanctions list.
- *Staking*: Validators who wish to earn additional rewards (through MEV[29]) decide whether the blocks they propose contain OFAC-compliant transactions or not. In general, OFAC violations can result in significant fines; however, the specific legal treatment on these matters has not been settled with regard to sanctions compliance and the technical mechanisms of block building and staking.
- *Exchanges*: Similar to the custody and trading use cases, exchanges need to screen clients and also ensure any transactions through their platform do not originate, or end up, in sanctioned wallets.
- *Tokenisation*: Tokenisation risks and mitigations are identical to the custody, trading and exchange use cases.

## PRACTICAL CHALLENGES IN APPLYING A BLOCKCHAIN RISK FRAMEWORK

This paper has described blockchain protocol risks as a foundational category of risks affecting all digital assets, and other products and services that involve the use of public blockchain systems. A comprehensive risk management framework was introduced, covering risk identification, measurement, implementation of controls, ongoing monitoring, and reporting, while also shedding light on the practical challenges and considerations involved in the process.

Unlike traditional risk management, which focuses on centrally controlled organisations or third-party risks in the use of software systems, public blockchain networks do not have a central authority providing risk assurances. Consequently, the onus of risk management falls upon the parties building or investing in the blockchain until a suitable certification or common methodology for decentralised standards is developed.

When implementing a risk management system for blockchain networks, institutions must weigh the choice between building their own solution or purchasing one from a reputable provider. Factors influencing this decision include the speed of implementation, alignment with industry standards and the prospect of ongoing innovation and collaboration with specialised vendors possessing deep expertise in the field.

The disruptive potential of blockchain technology presents numerous opportunities for financial institutions, but, concurrently, it intensifies the focus of regulators, legislators and policymakers on the risks involved. As the regulatory landscape evolves, institutions may be subject to more formal obligations to ensure responsible innovation (see recent guidelines from Basel Committee on Banking Supervision for addressing vulnerabilities in blockchain infrastructure[30]).

To fully capitalise on the opportunities offered by blockchain networks and digital assets, it is imperative for institutions to strike a balance between innovation and risk management. By proactively monitoring and mitigating risks, institutions can foster responsible innovation, harness the transformative power of blockchain technology, and ultimately create a secure and sustainable financial ecosystem.

## APPENDIX A OPERATIONALISING THE RISK MANAGEMENT FRAMEWORK

Figure A1, which is a screenshot of a blockchain risk management platform developed at Metrika,

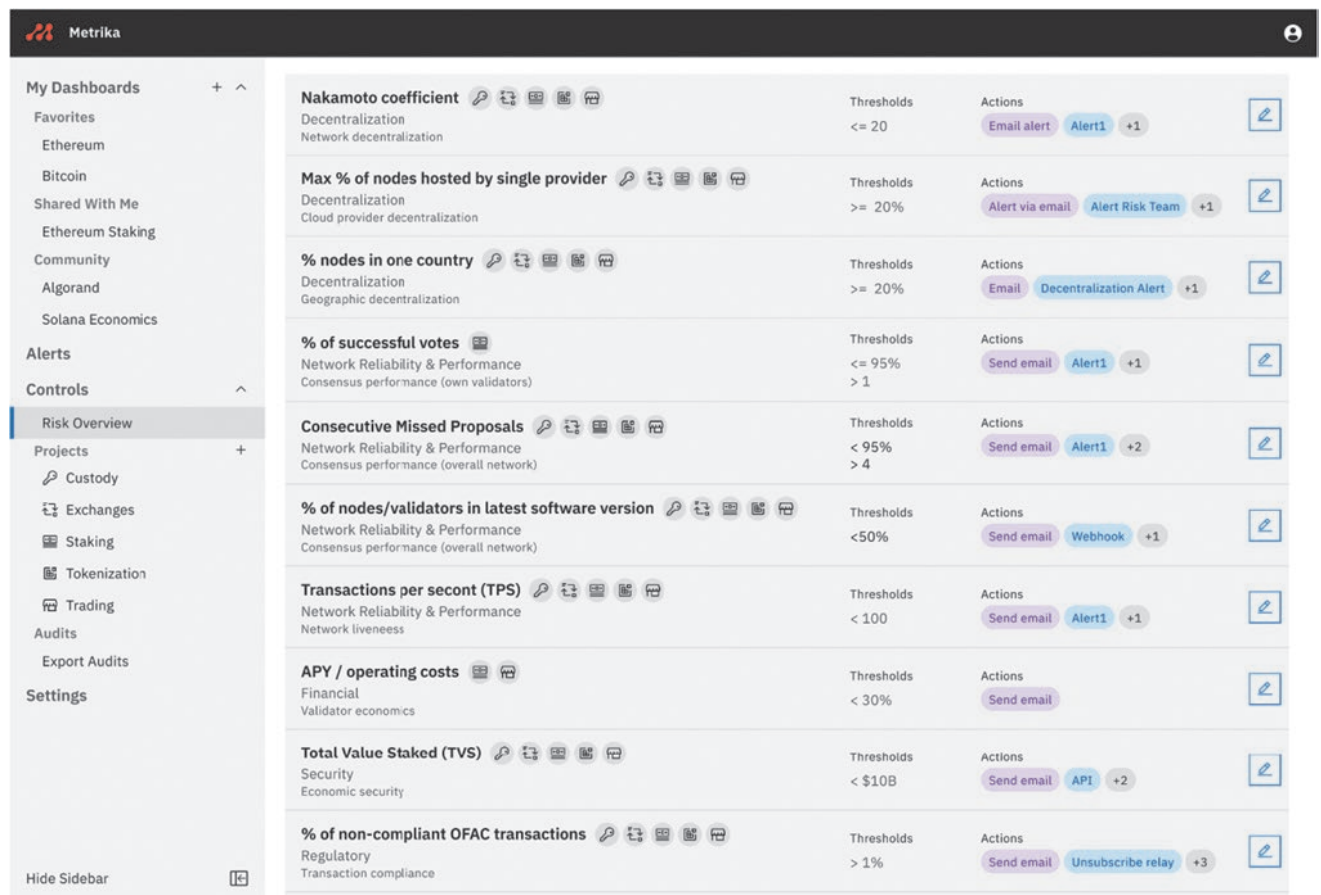

**Figure A1:** KRIs and alerting conditions defined by an institution using Metrika. KRIs and alerts are only for illustrative purposes

showcases how practitioners can effectively operationalise the proposed framework. The platform allows users to choose from an expanding list of KRIs, and tailor-based on the institution's internal risk appetites and thresholds. By doing so, institutions can leverage Metrika to define their risk management framework, monitor KRIs in real-time, react to issues faster and streamline reporting for internal and external audits and examinations (Figure A1).

## REAL-WORLD EXAMPLE OF MANAGING AND MONITORING KRIS

On 12th May, 2023, Ethereum Mainnet lost its ability to finalise blocks for a little over an hour. This was the second time over a period of 24 hours that Ethereum exhibited finality issues. As discussed in the paper, finality is a key property of blockchain networks, as it ensures that transactions will be persisted on chain permanently. Without finality, there are no such guarantees.

A KRI for managing and monitoring finality on Ethereum is the distance between the current epoch and the latest finalised epoch of its consensus layer, also known as Beacon Chain. During optimal operation, the distance between epochs is equal to 2, so, from a risk management standpoint, a crucial alert would trigger as soon as this distance became greater than 2. The chart on the left of Figure A2 shows the current and finalised epochs on 12th May, 2023, where, around 17:20 UTC Ethereum stopped finalising blocks. The previously mentioned alert should be followed by the appropriate response (eg halt trading, increase confirmation time for pending transactions), at the very least throughout the duration of the incident. Real-time monitoring during such times is imperative, as interventions may be deemed necessary to mitigate the risks associated with the incident. Furthermore, once the incident is over, the incident response team may need to reverse any actions taken to ensure business continuity.

As the post-mortem report of this particular incident indicated,[31] the root cause was that two consensus clients, namely Prysm and Teku, could not process old but valid attestations (ie votes) in an optimal way, leading to resource exhaustion. The chart on the right of Figure A2 shows the CPU usage of a validator running the Prysm consensus client before, during and after the finality incident — note that while not mentioned in the paper, validator CPU usage can be a KRI of its own. It is quite evident that, during normal operations, the average CPU usage was around 10 per cent, whereas it reached 90 per cent during the incident. Managing and monitoring multiple KRIs is crucial for assessing the severity of issues and troubleshooting.

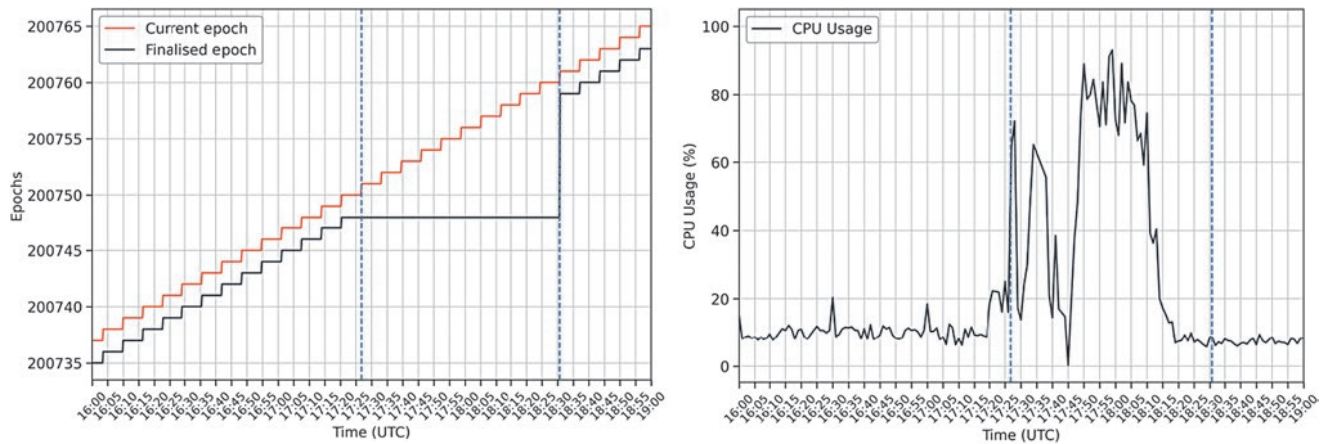

**Figure A2:** (Left) Current and finalised epochs from 16:00 UTC until 19:00 UTC on 12th May, 2023. The vertical blue dashed lines indicate the start and end of the incident; (Right) CPU Usage of an Ethereum Mainnet validator running the Prysm consensus client from 16:00 UTC until 19:00 UTC on 12th May, 2023

## References and notes

1 Hayes, A. (2023) 'Blockchain Facts: What Is It, How It Works, and How It Can Be Used', available at: https://www.investopedia.com/terms/b/blockchain.asp (accessed 27th September, 2022).

2 The terms blockchain protocol risk, blockchain risk and protocol risk are used interchangeably.

3 'Regulation (EU) 2022/2554 of the European Parliament and of the Council', available at: https://eur-lex.europa.eu/legal-content/EN/TXT/PDF/?uri=CELEX:32022R2554 (accessed May, 2022).

4 'Cryptocurrency Market Structure Draft Bill', available at: https://docs.house.gov/meetings/AG/AG00/20230606/116051/HHRG-118-AG00-20230606-SD003.pdf (accessed 20th June, 2023).

5 There are cases where smart contracts, eg those relating to bridges and oracles, can affect many other smart contracts.

6 Mishunin, D. (2018) 'UTXO and Account Model Comparison', available at: https://blog.hashex.org/utxo-and-account-model-comparison-c4098a9bc119 (accessed 5th April, 2023).

7 Blockchain (n.d.) 'Hashrate Distribution', available at: https://www.blockchain.com/explorer/charts/pools (accessed 1st April, 2023).

8 Hayes, A. (2008) 'Gini Index Explained and Gini Coefficients Around the World', available at: https://www.investopedia.com/terms/g/gini-index.asp (accessed 4th April, 2023).

9 United States Census Bureau (last revised 8th October, 2021) 'Theil Index', available at: https://www.census.gov/topics/income-poverty/income-inequality/about/metrics/theil-index.html (accessed 4th April, 2023).

10 Lutz, S. (2022) 'Is Solana Decentralized? Cloud Provider Hetzner Ban Raises Questions', available at: https://decrypt.co/113429/is-solana-decentralized-cloud-provider-hetzner-ban-raises-questions (accessed 1st April, 2023).

11 Quarmby, B. (2021) 'Solana Attributes Major Outage to Denial-of-Service Attack Targeting DEX Offering', available at: https://cointelegraph.com/news/solana-attributes-major-outage-to-denial-of-service-attack-targeting-dex-offering (accessed 1st April, 2023).

12 Reynolds, S. (2023) 'Avalanche Blockchain's X and C Networks See Brief Outage', available at: https://www.yahoo.com/entertainment/avalanche-blockchains-x-c-networks-073731219.html (accessed 1st April, 2023).

13 Hopper, G. (2022) 'DeFi: New Risks Require New Regulation', available at: https://www.financialresearch.gov/frac/files/OFR_FRAC_DeFi-Hopper.pdf (accessed 8th November, 2022).

14 United States District Court Southern District of New York (22nd December, 2020) 'Securities Exchange Commission, Plaintiff, against Ripple Labs, Inc., Bradley Garlinghouse and Christian A. Larsen, Defendants', available at: https://www.sec.gov/litigation/complaints/2020/comp-pr2020-338.pdf (accessed 1st April, 2023).

15 SEC (9th February, 2023) 'Kraken to Discontinue Unregistered Offer and Sale of Crypto Asset Staking-As-A-Service Program and Pay $30 Million to Settle SEC Charges', available at: https://www.sec.gov/news/press-release/2023-25 (accessed 1st April, 2023).

16 Metrika (5th January, 2023) 'Exploring the Impact of OFAC Compliance on MEV Relays: An Investigation', available at: https://blog.metrika.co/exploring-the-impact-of-ofac-compliance-on-mev-relays-an-investigation-80ef3def4446 (accessed 1st April, 2023).

17 It is certainly possible for organisations to devise and calibrate the first-order weights, although that makes the category-based weighting approach more cumbersome from an implementation standpoint.

18 BNY Mellon (11th October, 2022) 'BNY Mellon Launches New Digital Asset Custody Platform', available at: https://www.bnymellon.com/us/en/about-us/newsroom/press-release/bny-mellon-launches-new-digital-asset-custody-platform-130305.html (accessed 1st April, 2023).

19 BNY Mellon (n.d.) 'Migration to Digital Assets Accelerates: Digital Asset Survey', available at: https://www.bnymellon.com/us/en/insights/all-insights/digital-asset-survey.html (accessed 1st April, 2023).

20 Metrika (n.d.) 'Institutional Staking Guide', available at: https://reports.metrika.co/institutional-staking-whitepaper (accessed 1st April, 2023).

21 Pereira, A. P. (2022) 'Staking Providers Could Expand Institutional Presence in the Crypto Space: Report', available at: https://cointelegraph.com/news/staking-providers-could-expand-institutional-presence-in-the-crypto-space-report (accessed 1st April, 2023).

22 Rosenberg, E. (2022) 'What Are Cross-Chain Bridges?', available at: https://www.investopedia.com/what-are-cross-chain-bridges-6750848 (accessed 1st April, 2023).

23 Shen, M. (2020) 'OKEx Mulls ETC Delisting after Losses from Two 51 per cent Attacks', available at: https://www.yahoo.com/video/okex-mulls-etc-delisting-losses-194834208.html (accessed 3rd April, 2023).

24 Michaels, D. and Huang, V. G. (2023) 'SEC Sues Coinbase, Alleges It Is Unregistered Broker', available at: https://www.wsj.com/articles/sec-sues-coinbase-says-it-is-unregistered-broker-dealer-95ec0637l (accessed 31st May, 2023).

25 Nassr, I. K. (2020) 'The Tokenisation of Assets and Potential Implications for Financial Markets', available from https://www.oecd.org/finance/The-Tokenisation-of-Assets-and-Potential-Implications-for-Financial-Markets-HIGHLIGHTS.pdf (accessed 1st April, 2023).

26 Important note: The alerting conditions are only used for illustration purposes. Furthermore, thresholds should be adjusted specifically to each.

27 Yu, D. (2020) 'Coinbase's Perspective on the Recent Ethereum Classic (ETC) Double Spend Incidents', available at: https://www.coinbase.com/blog/coinbases-perspective-on-the-recent-ethereum-classic-etc-double-spend (accessed 5th April, 2023).

28 US Department of the Treasury (n.d.) 'U.S. Treasury Sanctions Notorious Virtual Currency Mixer Tornado Cash', available at: https://home.treasury.gov/news/press-releases/jy0916 (accessed 5th April, 2023).

29 Ethereum (last edit 1st June, 2023) 'Maximal Extractable Value (MEV)', available at: https://ethereum.org/en/developers/docs/mev/ (accessed 5th April, 2023).

30 Basel Committee on Banking Supervision (December 2022) 'Prudential Treatment of Cryptoasset Exposures', available at: https://www.bis.org/bcbs/publ/d545.pdf (accessed 5th April, 2023).

31 Das, N., Tsao, T., Van Loon, P., Potuz, K. K. and He, J. (11th May, 2023) 'Post-Mortem Report: Ethereum Mainnet Finality (05/11/2023)', available at: https://offchain.medium.com/post-mortem-report-ethereum-mainnet-finality-05-11-2023-95e271dfd8b2 (accessed 20th May, 2023).